\documentclass[aps,pra,twocolumn,showpacs,superscriptaddress,floatfix]{revtex4}
\usepackage{graphicx}
\usepackage{bm}
\begin{document}
\title{Magnetically tunable Feshbach resonances in ultracold Li-Yb mixtures}
\author{Daniel A. Brue}
\email{E-mail: daniel.brue@durham.ac.uk}%
\affiliation{Department of Chemistry, Durham University, South Road,
Durham, DH1~3LE, United Kingdom}
\author{Jeremy M. Hutson}
\email{E-mail: J.M.Hutson@durham.ac.uk}%
\affiliation{Department of Chemistry, Durham University, South Road,
Durham, DH1~3LE, United Kingdom}

\date{\today}

\begin{abstract}
We investigate the possibility of forming Li+Yb ultracold molecules by
magnetoassociation in mixtures of ultracold atoms. We find that
magnetically tunable Feshbach resonances exist, but are extremely
narrow for even-mass ytterbium isotopes, which all have zero spin. For
odd-mass Yb isotopes, however, there is a new mechanism due to
hyperfine coupling between the electron spin and the Yb nuclear
magnetic moment. This mechanism produces Feshbach resonances for
fermionic Yb isotopes that can be more than 2 orders of magnitude
larger than for the bosonic counterparts.
\end{abstract}


\maketitle

It has recently become possible to convert gases of alkali-metal atoms
coherently into molecules in their absolute ground state, by
magnetoassociation followed by stimulated Raman adiabatic passage
(STIRAP). This has so far been achieved for KRb \cite{Ni:KRb:2008} and
Cs$_2$ \cite{Danzl:ground:2010}, but other species are under active
investigation. Other methods for producing ultracold ground-state
molecules are also under development \cite{Sage:2005, Lang:ground:2008,
Viteau:2008, Deiglmayr:2008, Mark:2009, Haimberger:2009}. Ultracold
polar molecules have important potential applications in quantum
computing and quantum simulation, and dipolar quantum gases promise a
rich new physics with many novel properties \cite{Carr:NJPintro:2009}.

Even more new properties would be accessible for ultracold molecules
with both electric and magnetic dipole moments \cite{Micheli:2006}.
However, alkali-metal dimers have singlet ground states, which cannot
be tuned magnetically except through the very small magnetic moments of
the constituent nuclei. There is therefore great interest in producing
ultracold polar molecules with doublet or triplet ground states. Among
the most promising candidates for this are molecules formed from an
alkali-metal atom and a $^1S$ atom such as an alkaline earth or Yb. A
number of experimental groups are studying ultracold mixtures in which
such molecules might be formed, including Rb+Yb \cite{Nemitz:2009} and
Li+Yb \cite{Okano:2010, Ivanov:2011, Hara:2011}.

It was originally believed that systems of this type would not have
Feshbach resonances suitable for magnetoassociation, because the
bound-state channels (correlating with the upper hyperfine state of the
alkali-metal atom) are uncoupled to the entrance channel by the
conventional collision Hamiltonian. However, we have recently shown
that a coupling does exist, due to the modification of the alkali-metal
hyperfine coupling by the singlet atom at short range
\cite{Zuchowski:RbSr:2010}. For the prototype system Rb+Sr, we showed
that, under favorable circumstances, this mechanism could produce
magnetically tunable Feshbach resonances with widths up to 100 mG,
which might be used for magnetoassociation.

In the present paper we consider the topical system Li+Yb, which is
under experimental investigation by at least two groups and has also
been investigated using electronic structure calculations
\cite{Gopakumar:2010, Zhang:2011}. We show that the mechanism we
previously proposed does produce Feshbach resonances in this system,
but that for bosonic (spin-zero) isotopes of Yb they are extremely
narrow. However, for fermionic isotopes of Yb there is a different
mechanism that can produce resonances that are {\em more than 2 orders
of magnitude} wider. Furthermore, the recent measurements of the
scattering length for $^6$Li-$^{174}$Yb \cite{Ivanov:2011, Hara:2011}
allow us to estimate the magnetic fields at which these Feshbach
resonances will occur.

The collision Hamiltonian for a pair of atoms $a$ and $b$ is
\begin{equation}
\frac{\hbar^2}{2\mu}\left[-R^{-1}\frac{d^2}{dr^2}R +
\frac{\hat L^2}{r^2}\right] + \hat H_a + \hat H_b + \hat V(R),
\end{equation}
where $r$ is the internuclear distance, $\mu$ is the reduced mass,
$\hat L^2$ is the angular momentum operator for mechanical rotation of
the atoms about one another, $\hat H_a$ and $\hat H_b$ are the
Hamiltonians for the free atoms (in an applied field) and $\hat V(r)$
is the interaction operator. For collision of an alkali-metal atom $a$
with a closed-shell ($^1$S) atom $b$,
\begin{eqnarray}
\hat H_a &=& \zeta_a
\hat i_a \cdot \hat s + \left( g_a^e \mu_{\rm B} \hat s_z +
g_{a}^{\rm nuc} \mu_{\rm N} \hat i_{a,z}\right) B,
 \label{eq:hmon} \cr
\hat H_b &=& g_b^{\rm nuc} \mu_{\rm N} \hat i_{b,z} B,
\end{eqnarray}
where $\zeta_a$ is the hyperfine coupling constant for atom $a$, $\hat
s$, $\hat i_a$ and $\hat i_b$ are the electron and nuclear spin
operators, $g_a^e$, $g_a^{\rm nuc}$  and $g_b^{\rm nuc}$ are the
corresponding $g$-factors, and $B$ is the magnetic field, whose
direction defines the $z$-axis. There is only one electronic state, of
symmetry $^2\Sigma^+$, arising from interaction of atoms in $^2$S and
$^1$S states. As described in ref.\ \cite{Zuchowski:RbSr:2010}, the
hyperfine coupling constant of the alkali metal is modified by the
presence of the closed-shell atom, $\zeta_a(R) = \zeta_a +
\Delta\zeta_a(R)$. However, if the closed-shell atom has non-zero
nuclear spin $i_b$, there may also be a coupling
$\zeta_b(R)=\Delta\zeta_b(R)$ between $i_b$ and the electron spin; this
term has not been considered previously. The interaction operator is
therefore
\begin{equation}
\hat V(r) = V(R) + \Delta\zeta_a(R) \hat i_a \cdot \hat s +
\Delta\zeta_b(R) \hat i_b \cdot \hat s.
\label{eq:vint}
\end{equation}
There are additional small contributions to $\hat V(R)$ from nuclear
quadrupole coupling, electron-nuclear dipolar coupling and
spin-rotation interactions, but the $\Delta\zeta$ terms have by far the
largest effect \cite{Zuchowski:RbSr:2010}.

\begin{figure}
\includegraphics[width=\linewidth]{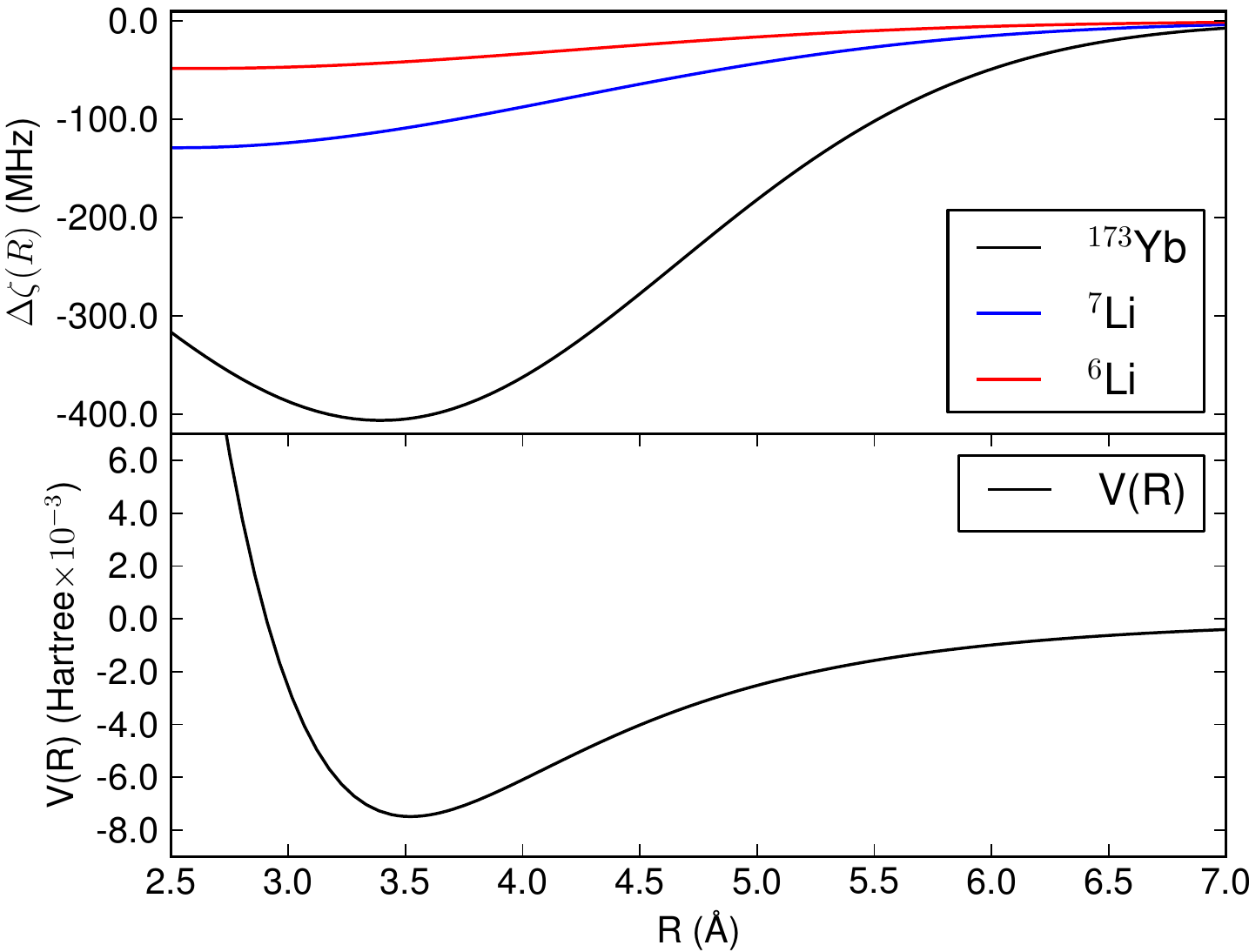}
\caption{Interaction potential $V(R)$ for LiYb (lower panel)
\cite{Zhang:2011} and R-dependence of hyperfine coupling constants
$\Delta\zeta(R)$ (upper panel).} \label{pots}
\end{figure}

In the present work, we obtained the quantities $\zeta_{\rm Li}(R)$ and
$\zeta_{\rm Yb}(R)$ with the relativistic density-functional theory
(DFT) approach \cite{vanLenthe:1994} implemented in the ADF program
\cite{ADF1}, using the PBE0 functional \cite{Adamo:1999} with a QZ4P
quadruple-zeta relativistic basis set \cite{VanLenthe:2003}. The
asymptotic value of $\zeta_{\rm Li}$ for $^7$Li was underestimated by
5.8\% in the DFT calculations, so we scaled $\zeta_{\rm Li}(R)$ to
reproduce the experimental atomic value. The resulting quantities
$\Delta\zeta(R)$ are shown in Figure \ref{pots}; they were fitted to
Gaussian forms $\zeta_0 e^{-a(r-r_c)^2}$, with parameters
$\zeta_0=-129$ MHz, $a=0.101$ \AA$^{-2}$ and $r_c=2.60$~\AA\ for
$^{7}$Li and $\zeta_0=-406$ MHz, $a=0.31$ \AA$^{-2}$ and
$r_c=3.39$~\AA\ for $^{173}$Yb. This corresponds to a 32\% maximum
reduction in $\zeta$ for ${}^7$Li (and also for ${}^6$Li). The values
for $^6$Li and $^{171}$Yb are obtained by scaling according to the
nuclear magnetic moments, giving $\zeta_0=-48$ MHz and $-1500$ MHz
respectively.

Zhang {\em et al.} \cite{Zhang:2011} have carried out detailed studies
of the potential curves of the ground and excited states of LiYb. In
the present work, we used their CCSD(T)/ECP potential points
\cite{Zhang:private:2011}, obtained from large-basis coupled-cluster
calculations using a fully relativistic effective core potential for Yb
and including counterpoise corrections. These were extrapolated at long
range using the form $V(r)=-C_6R^{-6}$, with coefficient $C_6=1594$
$E_{\rm h} a_0^6$ \cite{Zhang:2011}. The short- and long-range parts of
the potential were smoothly connected between 20 and 25 $a_0$ with the
switching function used by Janssen {\em et al.} \cite{Janssen:2009}.

For a potential with known long-range behavior, both the energies of
high-lying bound states \cite{LeRoy:1970} and the scattering length
\cite{Gribakin:1993} may be expressed in terms of the fractional part
of the quantum number at dissociation, or equivalently in terms of a
semiclassical phase integral, $\Phi = \int k(R) dR$, where $k^2(R)=2\mu
V(R)/\hbar^2$. The potential curve of Zhang {\em et al.}\
\cite{Zhang:2011} supports 24 bound states for $^6$LiYb. However, it is
probably accurate to only a few percent, and variations within this
uncertainty are enough to span the entire range of scattering length
$a$ from $+\infty$ to $-\infty$. However, Ivanov {\em et al.}
\cite{Ivanov:2011} and Hara {\em et al.} \cite{Hara:2011} have recently
measured values of the scattering length to be $|a|=13\pm3\ a_0$
($0.69\pm 0.16$ nm) and $1.0\pm0.2$~nm, respectively. Because the sign
of $a$ is unknown, we have scaled the potential of Zhang {\em et al.}\
to give two potentials, $V_+$ and $V_-$, with scattering lengths
$a=+0.8$ nm and $-0.8$ nm, each within the error bars of both
measurements.

\begin{figure}
\includegraphics[width=\linewidth]{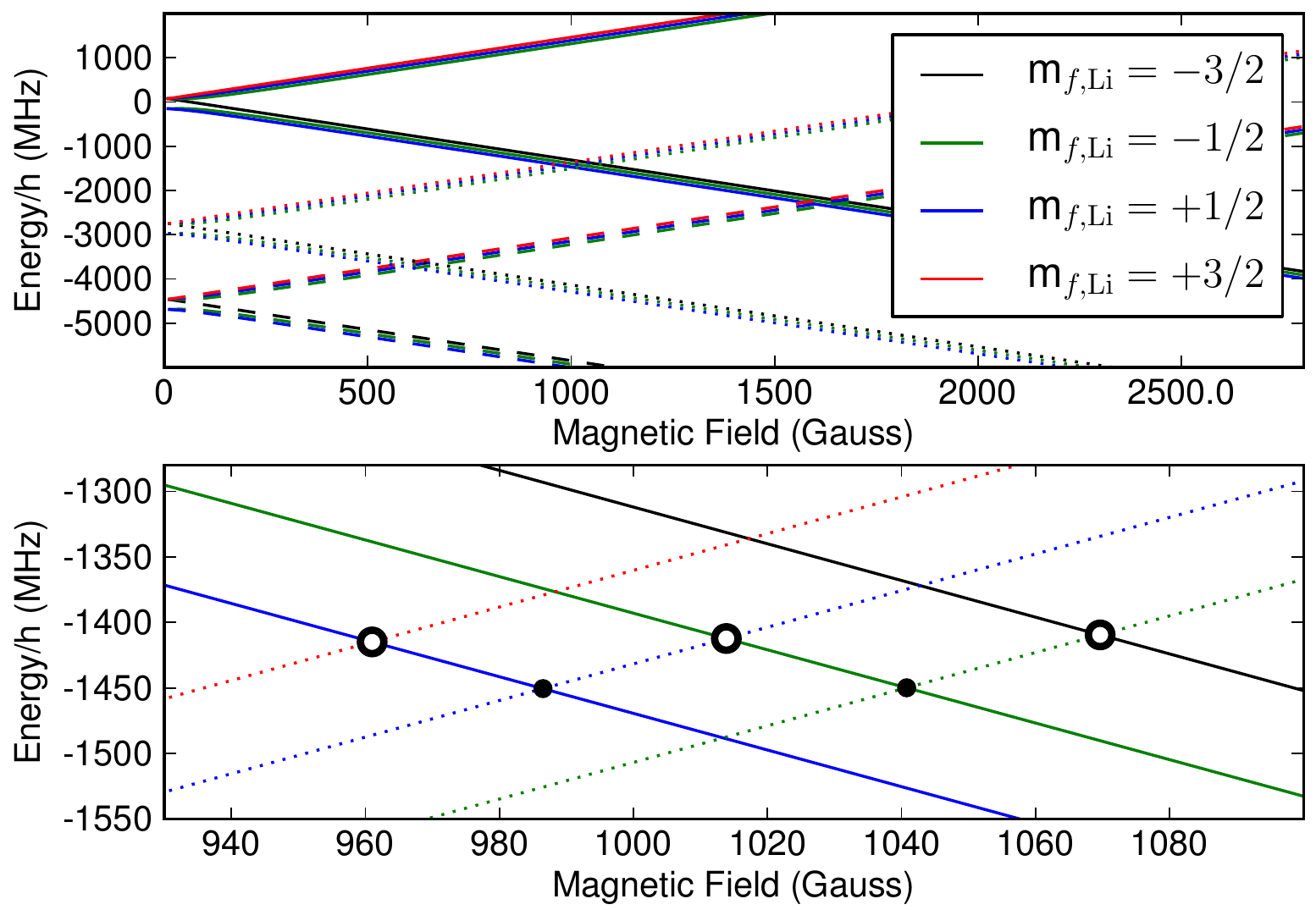}
\caption{(Color online) Upper panel: Molecular levels (dotted and
dashed) that cross atomic thresholds (solid) for $^{6}$LiYb, calculated
for potentials $V_+$ (dotted lines) and $V_-$ (dashed lines) as a
function of magnetic field. The Yb isotope chosen makes little
difference on the scale of the Figure. The lower panel shows an
expanded view of the crossing region for potential $V_+$. Filled circles
show positions where narrow resonances occur for all isotopes and open circles
show where wider resonances occur for Yb isotopes with non-zero spin.}
\label{thresh}
\end{figure}

Zero-energy Feshbach resonances of the type that might be suitable for
magnetoassociation occur at magnetic fields where molecular levels
cross atomic thresholds. In a molecule such as LiYb, the entire
molecular Hamiltonian is nearly diagonal in a basis set of
field-dressed atomic functions, so that the molecular levels lie
parallel to the atomic thresholds as a function of magnetic field. This
is shown in Fig.\ \ref{thresh}: for the $V_+$ potential, there is a
group of crossings around 1000~G that might in principle cause Feshbach
resonances, whereas for $V_-$ the corresponding crossings occur around
1600~G.

In the present work, we characterize resonances by carrying out
coupled-channel scattering calculations using the MOLSCAT program
\cite{molscat:v14}, modified to handle collisions of atoms in magnetic
fields \cite{Gonzalez-Martinez:2007, Hutson:Cs2:2008}. The molecular
wavefunctions are conveniently expanded in an uncoupled basis set
$|sm_s\rangle |i_{\rm Li} m_{i,{\rm Li}}\rangle |i_{\rm Yb} m_{i,{\rm
Yb}}\rangle|L M_L\rangle$. The Hamiltonian is diagonal in the total
projection quantum number $M_{\rm tot}=M_F+M_L$, where
$M_F=m_s+m_{i,{\rm Li}}+m_{i,{\rm Yb}}$. This provides the scattering
length $a(B)$, which (in the absence of inelastic scattering
\cite{Hutson:res:2007}) follows the functional form $a(B)=a_{\rm
bg}\left[1 + \Delta / (B - B_{\rm res})\right]$. The width $\Delta$ is
conveniently obtained from the difference between the positions of the
pole and zero in $a(B)$. In the present work, we extended MOLSCAT to
provide an option to {\em converge} on poles and zeroes of $a(B)$,
instead of extracting them from a fit to a grid of points.

There are only very small terms in the collision Hamiltonian that
couple states of different $L$ or $M_L$. For mixtures containing
spin-zero isotopes of Yb (mass numbers 168, 170, 172, 174 and 176),
conservation of $M_{\rm tot}$ and $M_L$ requires that the Li projection
quantum number $m_{f,{\rm Li}}=m_s+m_{i,{\rm Li}}$ is also conserved.
Because of this, Feshbach resonances can occur only where states of the
LiYb molecule cross atomic thresholds with the same value of $m_{f,{\rm
Li}}$, as shown by solid circles in Fig.\ \ref{thresh}. The
lowest-field resonances for the $V_+$ and $V_-$ potentials for
$^6$Li$^{174}$Yb are shown in Table \ref{tab:res}. However, in both
cases the resonances are extremely narrow ($\Delta < 10$ $\mu$G) and
would be very challenging to observe experimentally. This arises mostly
because the $^6$Li hyperfine coupling constant is only 152~MHz in the
free atom, so that even a 32\% modification of it causes only very weak
coupling.

\begin{table}

\caption{ Calculated properties of LiYb Feshbach resonances caused by
low-field-seeking bound states ($m_s^{\rm bound}=+\frac{1}{2}$)
crossing the lowest $m_s=-1/2$ threshold state. The bound states are
labeled by the total angular momentum projection of the lithium atom
alone, $m_{f,{\rm Li}}^{\rm bound}$. The results for Li$^{174}$Yb are
for $M_{\rm tot}=m_{f,{\rm Li}}$ and those for $^6$Li$^{173}$Yb and
$^7$Li$^{173}$Yb are for $M_{\rm tot}=1$ and $+3/2$, respectively.}
\label{tab:res}
\begin{ruledtabular}
\begin{tabular}{lccrl}
 system          &  $B_{\rm res}$ (G) &  $a_{\rm bg}$ (nm) & $\Delta$ ($\mu$G)
& $m_{f,{\rm Li}}^{\rm bound}$  \\ \hline
&& $V_+$ potential &&  \\ \hline
$^6$Li$^{174}$Yb &  996  &   +0.80  &   1.89  & $+1/2$   \\
$^7$Li$^{174}$Yb &  261  &   +1.74  &  1.64  & $+1$     \\
$^6$Li$^{173}$Yb &  961  &   +0.82  &  824  & $+3/2$   \\
                 &  986  &   +0.82  &  1.03  & $+1/2$   \\
                 &  1013  &   +0.82  &  $<$0.1  & $-1/2$   \\
$^7$Li$^{173}$Yb &  213  &   +1.80  &   178     & $+2$     \\
                 &  253  &   +1.80  &   38.5     & +1       \\
                 &  307  &   +1.80  &  1.17   & \ \ 0    \\
                 &  388  &   +1.80  &  $<$0.1 & $-1$     \\
\hline
&& $V_-$ potential &&  \\ \hline
$^6$Li$^{174}$Yb &  1648 &  $-0.80$ & $-6.13$  &  +1/2    \\
$^7$Li$^{174}$Yb &   679 &   +0.73  &  14.9    &  +1      \\
$^6$Li$^{173}$Yb & 1608  &  $-0.76$ & $-2790$   &  $+3/2$  \\
                 & 1634  &  $-0.76$ & $-4.59$ &  $+1/2$  \\
                 & 1661  &  $-0.76$ & $<\left|-0.1\right|$ &  $-1/2$  \\
$^7$Li$^{173}$Yb &  615  &   +0.76  & 857       &  +2      \\
                 &  669  &   +0.76  &  31.1     &  +1      \\
                 &  733  &   +0.76  &  0.52     & \ \ 0    \\
                 &  810  &   +0.76  & $<$0.1    &  $-1$    \\
\end{tabular}
\end{ruledtabular}
\end{table}

The resonance positions for $^6$LiYb depend only very weakly on the Yb
isotope, because changing the Yb isotopic mass has very little effect
on the reduced mass for collisions with a light atom such as Li. The
$^6$LiYb resonance positions are also quite insensitive to the quality
of the potential curve. They do depend somewhat on the value of $C_6$:
for example, increasing $C_6$ by 1\% shifts the calculated resonance
positions to higher field by 0.5~G for fixed $a$. However, the main
source of uncertainty in the $^6$LiYb resonance positions comes from
the uncertainty in the measured scattering length for $^6$Li$^{174}$Yb,
and hence on the potential scaling, as described below.

Table \ref{tab:res} also includes predictions for $^7$LiYb, where the
resonances are a little wider because of the larger value of $\zeta_0$.
However, in this case the 17\% mass scaling increases the number of
bound states from 24 to 26: the scattering length obtained for
$^7$LiYb, and the positions of the resonances, are far more sensitive
to the depth of the potential. For $^7$LiYb, the predicted resonance
positions could easily be in error by several hundred Gauss. An
experimental measurement of the scattering length for $^7$LiYb would be
needed to pin this down more accurately. Nevertheless, it is clear from
Table \ref{tab:res} that the Feshbach resonances are very narrow for
mixtures of $^7$Li with even-mass Yb isotopes, and this conclusion will
not be altered by changes in the potential curve.

There is however a different mechanism that can create Feshbach
resonances for Yb isotopes with non-zero spin ($^{171}$Yb, $i=1/2$ and
$^{173}$Yb, $i=5/2$). Figure \ref{pots} shows that, as Li approaches
Yb, it polarizes the spin density of Yb and creates a hyperfine
coupling between the molecular spin and the nuclear magnetic moment of
Yb. The hyperfine coupling produced in this way is substantially larger
than the change in the Li hyperfine coupling. The operator
$\Delta\zeta_{\rm Yb}(R) \hat i_{\rm Yb} \cdot \hat s$ has matrix
elements that connect basis functions with $\Delta m_s=\pm1$ and
$\Delta m_{i,{\rm Yb}}=\mp1$. This coupling creates Feshbach resonances
where states of the LiYb molecule cross atomic thresholds with values
of $m_{f,{\rm Li}}$ that differ by $\pm1$, as shown by open circles in
Fig.\ \ref{thresh}.

We have further extended MOLSCAT to handle the additional basis sets
and couplings required for the $\Delta\zeta_{\rm Yb}(R)$ coupling term,
and have carried out coupled-channel calculations of the resonance
positions and widths for $^6$Li$^{173}$Yb and $^7$Li$^{173}$Yb. The
results are given in Table \ref{tab:res}. It may be seen that there are
now resonances {\em more than 2 orders of magnitude wider} than for
Li$^{174}$Yb, driven by a direct coupling involving $\Delta\zeta_{\rm
Yb}(r) \hat i_{\rm Yb} \cdot \hat s$. The calculated widths range from
0.8 to 2.8 mG in width for $^6$Li$^{173}$Yb.

The criterion for producing molecules at a narrow resonance is that the
field must vary smoothly enough to achieve adiabatic passage. Field
calibration and field inhomogeneity are not necessarily a problem; it
is high-frequency field noise that must be reduced to well below the
resonance width.

\begin{figure}
\includegraphics[width=\linewidth]{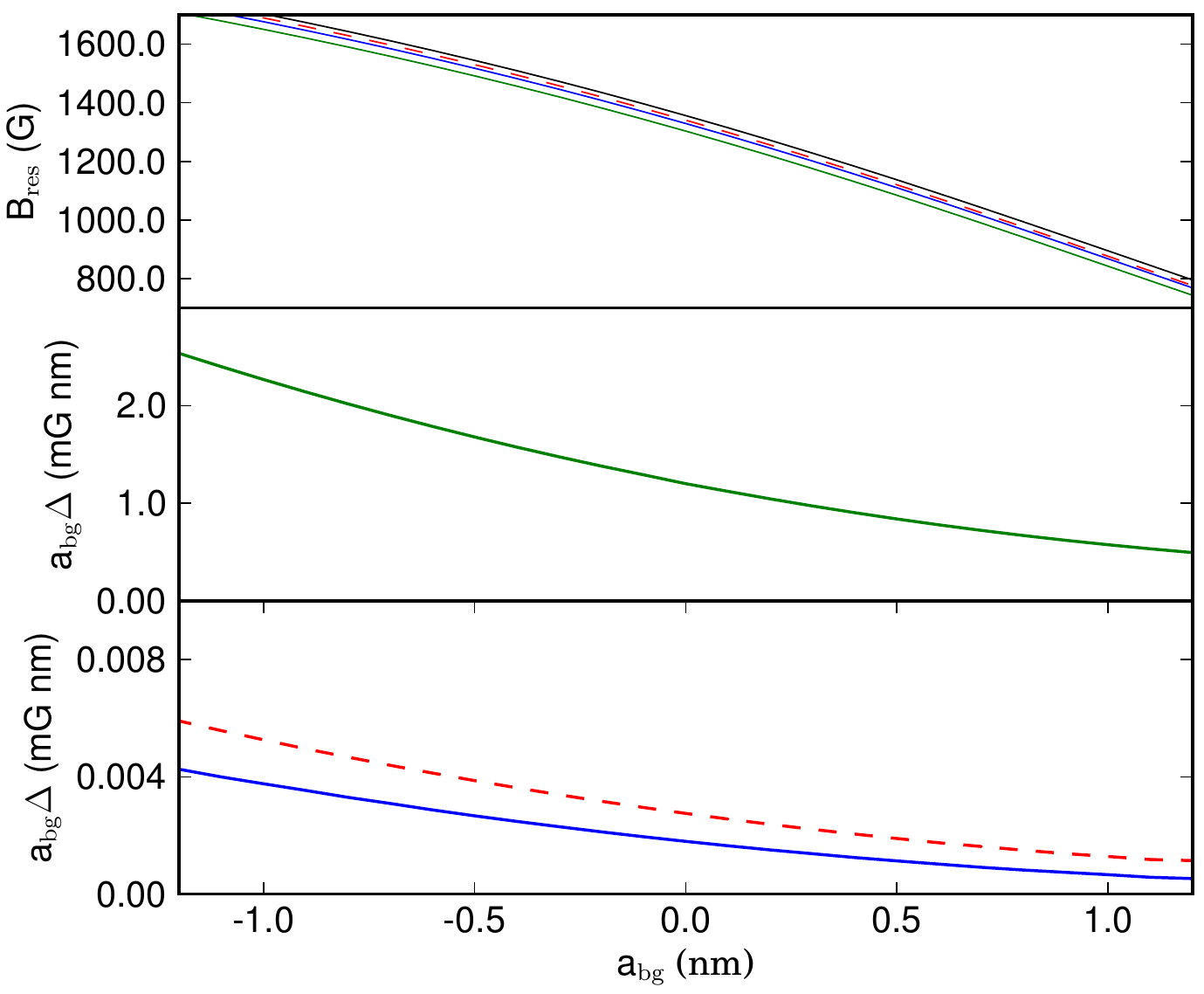}
\caption{(Color online) Positions and widths of the resonances at the
lowest $M_{\rm tot}$($m_s=-1/2$) threshold for $^{6}$Li$^{174}$Yb
(dashed) and $^{6}$Li$^{173}$Yb
(solid), as a function of the background scattering length produced by
scaling the potential. The width is not shown for the very narrow
$m_{f,{\rm Li}}^{\rm bound}=-3/2$ resonance. The widths are shown as $a_{\rm
bg}\Delta$ because the product varies smoothly whereas $\Delta$ itself
has a pole at $a_{\rm bg}=0.$} \label{6Li-track}
\end{figure}

The main uncertainty in the resonance positions for $^6$LiYb comes from
the uncertainty in the measured scattering length. Fig.\
\ref{6Li-track} shows how the resonance positions and widths for
$^6$Li$^{174}$Yb and $^6$Li$^{173}$Yb vary with the scattering length
used to scale the potential. For $^7$Li$^{173}$Yb, as for
$^7$Li$^{174}$Yb, the resonance positions depend quite strongly on the
quality of the potential curve. However, the resonance widths for all
isotopes depend mainly on the $\Delta\zeta(r)$ functions, obtained here
from DFT calculations as described above, and are fairly insensitive to
the depth of the potential.

At the fields where resonances occur for $^6$Li$^{173}$Yb, $m_{s}$ and
$m_{i,{\rm Yb}}$ are both reasonably good quantum numbers. For
$^6$Li$^{173}$Yb, each line in Figure \ref{6Li-track} represents
several levels with different values of $m_{i,{\rm Yb}}$ and thus
$M_{\rm tot}$. These produce resonances very close together (within 0.2
G), but with different widths, because the matrix element of $\hat
i_{\rm Yb} \cdot \hat s$ between the spin functions
$(m_s=+1/2,m_{i,{\rm Yb}})$ and $(m_s=-1/2,m_{i,{\rm Yb}}+1)$ is
$\frac{1}{2}[i_{\rm Yb}(i_{\rm Yb}+1)-m_{i,{\rm Yb}}(m_{i,{\rm
Yb}}+1)]^{1/2}$. The resonance widths are proportional to the square of
this. An experiment is likely to use Yb trapped in a single magnetic
sublevel and would therefore sample only one of these resonances; for
the lowest threshold ($m_{f,{\rm Li}}=+1/2$), the different possible
values for $m_{i,{\rm Yb}}$ of $-5/2$, $-3/2$, $-1/2$, $+1/2$ and
$+3/2$ lead to ratios in the resonance widths of 5:8:9:8:5,
respectively. These ratios are confirmed by the numerical results. The
resonance widths given for $^6$Li$^{173}$Yb in Table I, for $M_{\rm
tot}=0$, are for the widest of the group. There is no wide (Yb-driven)
resonance at this threshold for $m_{i,{\rm Yb}}=+5/2$. The Yb-driven
resonances for $^{171}$Yb are substantially narrower than those for
$^{173}$Yb, partly because $^{171}$Yb has a smaller nuclear magnetic
moment but mostly because the matrix element of $\hat i_{\rm Yb} \cdot
\hat s$ is much smaller (1 on the scale of the ratios above).

Resonances can also occur for Li atoms in magnetically excited states
($m_{f,{\rm Li}}=-1/2$ and $-3/2$) at the crossing points shown in
Fig.\ \ref{thresh}. However, the molecular states that are produced at
these resonances are quasibound at energies above the lowest threshold.
We have characterized these states by carrying out scattering
calculations immediately below the $m_{f,{\rm Li}}=-1/2$ threshold and
fitting them to obtain Breit-Wigner widths $\Gamma$ \cite{Ashton:1983}.
For the $V_+$ potential, the states with $m_{f,{\rm Li}}^{\rm
bound}=+3/2$ and $+1/2$ have calculated widths $\Gamma/h=715$ and 2.9
Hz, corresponding to lifetimes $\tau=\hbar/\Gamma=0.22$ and 55 ms
respectively, while the state with $m_{f,{\rm Li}}^{\rm bound}=-1/2$ is
too narrow to characterise. At decaying resonances of this type, the
real part $\alpha(B)$ of the complex scattering length $a(B)$ exhibits
an oscillation of magnitude $a_{\rm res}$ rather than an actual pole
\cite{Hutson:res:2007}. We obtain $a_{\rm res}=3.4$ fm, $1.3$ $\mu$m
and 4.0 $\mu$m, respectively, for these three resonances. For the two
higher-field resonances the behavior is sufficiently pole-like to
extract values of $\Delta$, which are 824 and 1.2 $\mu$G, respectively.

In conclusion, we have investigated the possibility of using Feshbach
resonances to produce ultracold LiYb molecules by magnetoassociation of
ultracold atoms. The mechanism that we previously identified for RbSr
\cite{Zuchowski:RbSr:2010} exists for LiYb, but produces only extremely
narrow Feshbach resonances in this case. However, we have identified a
new mechanism, which occurs only for Yb isotopes with non-zero nuclear
spin ($^{171}$Yb, $^{173}$Yb) that can produce Feshbach resonances that
are orders of magnitude wider. However, even for $^6$Li$^{173}$Yb the
largest calculated resonance width is 2.8 mG. Achieving the field
stability to produce molecules at such a narrow resonance will be
experimentally challenging.

\bibliography{../../all}

\end{document}